\documentclass{caosp309}
\usepackage{graphicx}
\usepackage{amsmath}
\usepackage{natbib}
\usepackage{float}
\articleNo{XX}
\pubyear{2024}
\volume{XX}
\volnumber{X}
\firstpage{X}
\received{13 Oct, 2024}
\accepted{28 Jan, 2024}

\def\BibTeX{{\rm B\kern-.05em{\sc i\kern-.025em b}\kern-.08em
             T\kern-.1667em\lower.7ex\hbox{E}\kern-.125emX}}
\begin{document}

\hauthor{S.\, Ta\c{s}demir and D. C. \c{C}ınar}

\title{A Detailed Analysis of Close Binary OCs}

\author{
        S.\, Ta\c{s}demir \inst{1}\orcid{0000-0003-1339-9148}
      \and
        D. C.\, \c{C}ınar \inst{1} \orcid{0000-0001-7940-3731}
}

\institute{$^1$Istanbul University, Institute of Graduate Studies in Science, Programme of Astronomy and Space Sciences, 34116, Beyaz{\i}t, Istanbul, Turkey.\\
           \email{tasdemirr.seval@gmail.com}
          }
\date{October 13, 2024}

\maketitle

\begin{abstract}
In this study, we analysed the close binary open clusters CWNU2666 and HSC224. They are in close spatial proximity, using photometric and astrometric data from the {\it Gaia} DR3. Most likely member stars were used with a membership probability ($P \geq 0.5$) were identified 106 and 146, the mean proper motion components ($\mu_{\alpha}\cos\delta$, $\mu_{\delta}$) obtained as  (0.646$\pm$0.155, -0.769$\pm$0.124) and (0.665$\pm$0.131, -0.728$\pm$0.107) mas yr$^{-1}$,  isochrone distances ($d_{\rm iso}$) as 1885$\pm$44 and 1866$\pm$29 pc, and ages ($t$) as 160$\pm$15 and 140$\pm$15 Myr, for CWNU 2666 and HSC 224, respectively.

\end{abstract}

\section{Introduction}
Open clusters (OCs) are stellar groups of hundreds or thousands of young stars that are bound together by the weak gravitational forces. Formed from the same gas and dust cloud, OCs typically share age and chemical properties, making them valuable tools for studying stellar evolution and Galactic structure and dynamics \citep{Friel_1995}. Recent observations have shown that some OCs are located quite close to each other and can interact gravitationally \citep{Song2022, Li2024, Haroon2024, Palma_2024}. Such clusters are called close binary open clusters (CBOCs). CBOCs provide unique opportunities to study their dynamical evolution and star formation in the Milky Way, serving as valuable laboratories for investigating both astrophysical processes and cluster interactions with the Galactic structure.

\section{Astrometric and Photometric Data}
\begin{figure}
\centering
\includegraphics[width=0.99\linewidth]{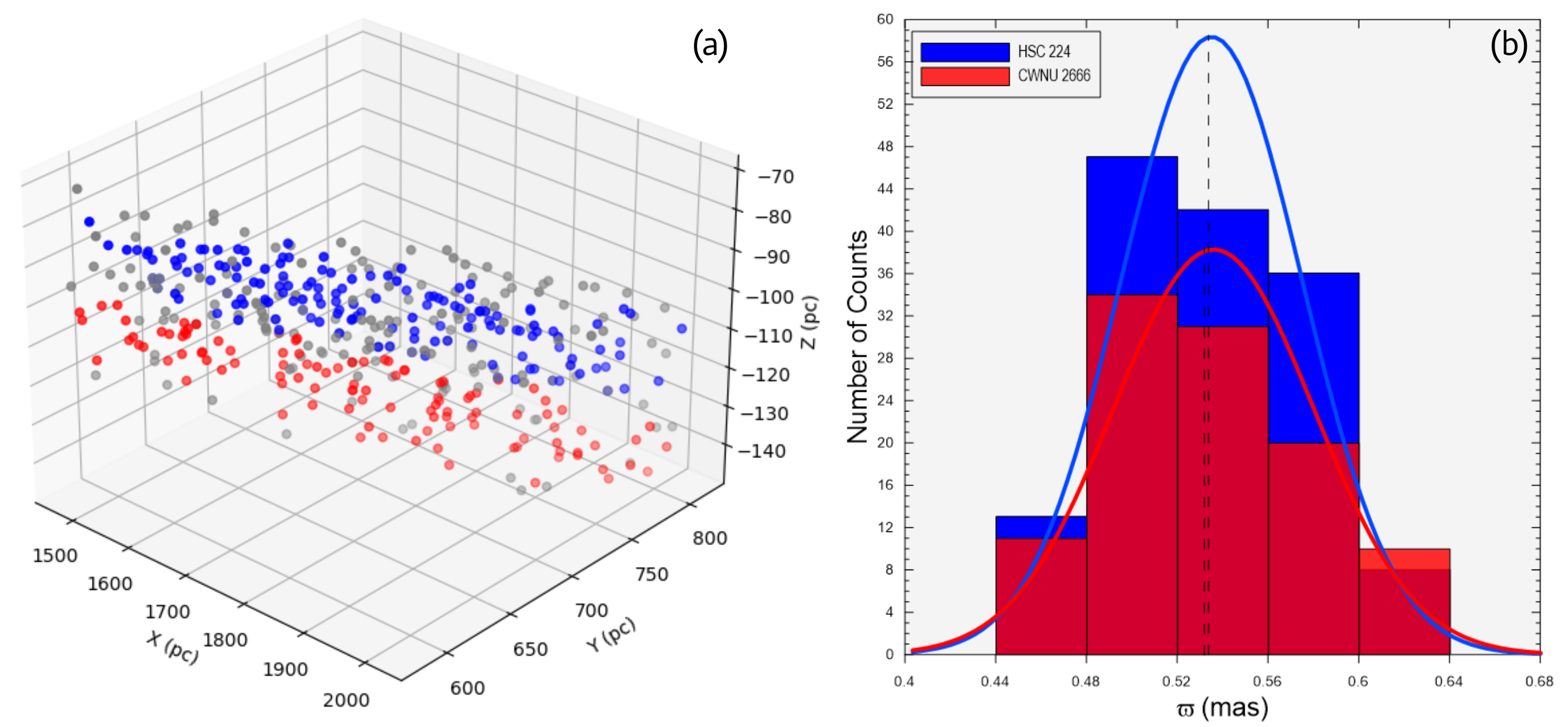}
\caption{Left panel (a) shows the 3D distribution of most likely stars ({\it P} $\geq$ 0.5) of HSC 224 (blue) and CWNU 2666 (red). Also grey circles represent the stars that are out of the limit radii of CBOCs. Right panel (b) illustrates the $Gaia$ DR3-based trigonometric parallax histogram. The black dashed lines represent the median of the Gaussian curves fitted to the parallax histograms for both clusters.}
\label{fig:Fig01}
\end{figure}

Tha photometric and astrometric anaylses of CWNU 2666 and HSC 224 utilized data from the \textit{Gaia} DR3 catalogued \citep[DR3;][]{Gaia_DR3}. The data were generated based on the equatorial coordinates provided by \citet{Hunt_2024} ($\alpha$, $\delta$) = (18$^{\rm h}$44$^{\rm m}$05$^{\rm s}$.33, -11$^{\rm o}$54${\rm '}$21${\rm ''}$.78). Figure~\ref{fig:Fig01}a presents the star field  for CWNU 2666 and HSC 224 in the $120\arcmin \times 120 \arcmin$ area.

\section{Analyses and Results}
This study provides a photometric and astrometric analysis of the CBOCs CWNU 2666 and HSC 224, as catalogued by \citet{Hunt_2024}. Using $Gaia$ DR3 data, we identified and differentiated member stars to determine their fundamental astrophysical and astrometric parameters \citep{Yontan_2022, Dursun_2024}.

\subsection{Spatial Distribution}

In this study, a spatial separation was performed by considering the central coordinates of the CBOCs. To accurately identify the most probable members of the cluster from field stars, we employed the Unsupervised Photometric Membership Assignment in Stellar Clusters ({\sc UPMASK}) method \citep{Martins_2014}, which utilizes astrometric parameters from $Gaia$ DR3 \citep{Tasdemir_2023}. This machine-learning approach, based on $k$-means clustering, resulting in the classification of 146 and 106 stars as probable ($P \geq 0.5$) members of CWNU 2666 and HSC 224, respectively and shown Figure~\ref{fig:Fig01}a. A histogram of the most probable members's trigonometric parallaxes was fitted with a Gaussian to estimate the mean, as shown in Figure~\ref{fig:Fig01}b. 

The 3D spatial positions of member stars in the open clusters CWNU 2666 and HSC 224 are analysed in heliocentric Cartesian coordinates ($X, Y, Z$) using the following transformations:
\begin{equation}
\begin{split}
X~&=~d~\cos\delta~\cos\alpha-R_\odot,\\
Y~&=~d~\cos{\delta}~\sin{\alpha},\\
Z~&=~d~\sin{\delta}.
\end{split}
\end{equation}
where $R_\odot$ = 8 kpc is the distance of Sun to the Galactic centre \citep{Majewski_1993}, $d$ is the distance, $\alpha$ and $\delta$ represent the equatorial coordinates, and the calculations are performed using data from high-probability member stars identified in the {\it Gaia} DR3 catalogue.

\begin{figure}
\centering
\includegraphics[width=0.90\linewidth]{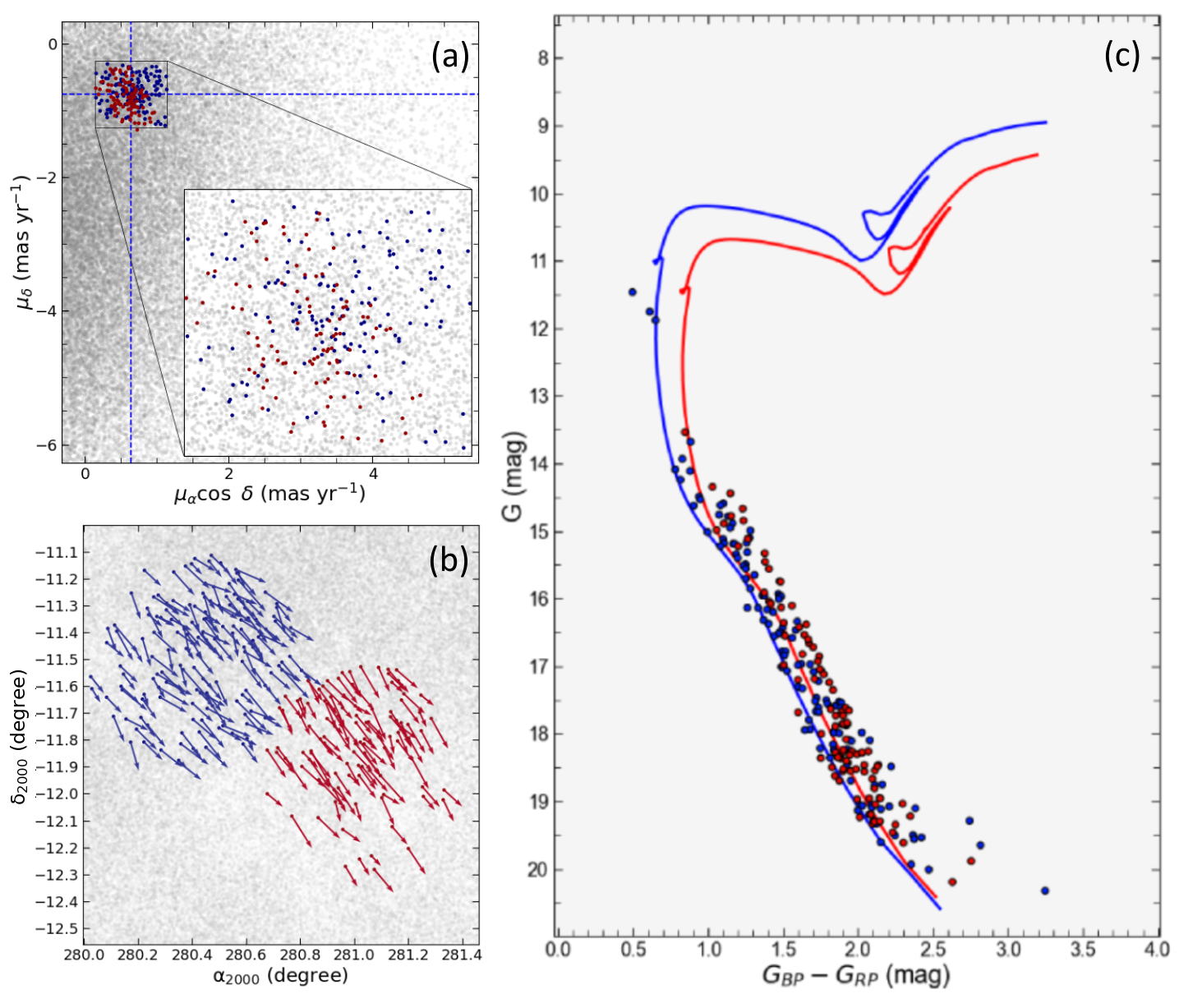}
\caption{Left-upper panel (a) the zoomed-box reveal regions with a high concentration member stars in the VPDs, left-bottom panel (b) proper-motion velocity vectors, and right panel (c) for the CWNU 2666 and HSC 224 in CMD.}
\label{fig:fig02}
\end{figure}

\subsection{Astrometric and Astrophysical Parameters}

To determine the mean proper-motion components of the clusters, we created the vector point diagram (VPD) from the most likely members and shown Figures \ref{fig:fig02}a and \ref{fig:fig02}b. {\sc UPMASK} method was employed for the membership analyses of the open clusters CWNU 2666 and HSC 224. This method utilised astrometric parameters, including proper motion components ($\mu_{\alpha}\cos \delta$, $\mu_{\delta}$) and trigonometric parallaxes ($\varpi$), along with their associated uncertainties, obtained from the {\it Gaia} DR3 catalogue. Consequently, the mean trigonometric parallaxes ($\varpi$) were obtained and via the linear equation, $d({\rm pc})=1000/\varpi$ (mas), converted to the distance ($d_{\varpi}$). The results are in a good agreement with \cite{Li2024}. 

For each of the two open clusters under study, we determined their astrophysical parameters using the isochrone-fitting method applied to high-probability member stars. This approach simultaneously estimates the colour excess, distance, and age of the clusters by fitting PARSEC isochrones to observational data on {\it Gaia}-based colour-magnitude diagrams. These results are compatible with \cite{Hunt_2024} listed in Table \ref{Tab01}. We utilized a classical technique for determining the fundamental astrophysical parameters of CBOCs called the main-sequence fitting method as shown in Figure \ref{fig:fig02}c \citep{Yontan_2023, Yontan-Canbay_2023, Elsanhoury_2024}.

\begin{table}
\footnotesize
\centering
\setlength{\tabcolsep}{3pt}
\caption{Astrometric and Astrophysical Parameters of CWNU 2666 and HSC 224 with comparisons to other published values.}
\begin{tabular}{lcc|cc}
\hline
& \multicolumn{2}{c}{This Study} & \multicolumn{2}{c}{\citet{Hunt_2024}}\\
Parameter & CWNU 2666 & HSC 224 & CWNU 2666 & HSC 224 \\
\hline
& \multicolumn{4}{c}{Astrometric Parameters} \\
\hline
$(\alpha, \delta)_{J2000}$ (Decimal) & \multicolumn{4}{c}{(281.02, -11.90) (280.23, -11.80)}\\
$(l, b)_{J2000}$ (Decimal) & \multicolumn{4}{c}{(21.49, -03.79) (21.23, -03.06)}\\
Members ($P \geq 0.5$) & 106 & 146 & 51 & 61\\
$\mu_{\alpha} \cos \delta$ (mas yr$^{-1}$) & $0.646\pm0.155$ & $0.665\pm0.131$ & 0.607$\pm$0.067 & 0.662$\pm$0.150\\
$\mu_{\delta}$ (mas yr$^{-1}$) & -$0.769\pm0.124$ & -$0.728\pm0.107$ & 	-0.803$\pm$0.100 &  -0.761$\pm$0.126\\
$\varpi$ (mas) & $0.537\pm0.006$ & $0.530\pm0.005$ & 0.563 $\pm$0.004 &0.552$\pm$0.004\\
$d_{\varpi}$ (pc) & $1863\pm102$ & $1857\pm169$ & 1776$\pm1$42 & 1811$\pm$138 \\
\textit{X, Y, Z  }(pc) & 1730, 681, -123 & 1729, 671, -99 & -6560, 614, -94 &-6547, 611, -73\\
$R_{\rm gc}$ (kpc) & 6.71 & 6.71 & -- & -- \\
\textit{k-means}& 25 & 34 & -- & -- \\
\hline
&\multicolumn{4}{c}{Astrophysical Parameters} \\
\hline
$E(G_{\rm BP}$-$G_{\rm RP})$ (mag) & $0.975\pm0.025$ & $0.807\pm0.016$ & 1.159$\pm$0.085 & 0.884$\pm$0.085\\
$A_{\rm G}$ (mag) & $1.816\pm0.047$ & $1.503\pm0.030$ & 2.159$\pm$0.158 &1.647$\pm$0.158\\
$[$Fe/H] (dex) & $0.010\pm0.05$ & $0.010\pm0.05$ & --&--\\
Age (Myr) & $160\pm15$ & $140\pm15$ & 98$\pm$43& 63$\pm$27\\
$G$ - $M_{\rm G}$ (mag) & $13.192\pm0.051$ & $12.857\pm0.034$ &-- &--\\
$d_{\rm iso}$ (pc) & $1885\pm44$ & $1866\pm29$ &1682$\pm$11 &1691$\pm$11\\
\hline
\label{Tab01}
\end{tabular}
\end{table}
\section{Future Works}
Future research will involve conducting spectral energy distribution (SED) analyses for the most probable member stars of each cluster, followed by a comparison with model atmosphere parameters from existing spectral studies in the literature. Additionally, the total masses of the clusters will be calculated for comparative purposes. These analyses aim to enhance our understanding of the previous interactions and formation processes of the clusters in relation to one another.

\acknowledgements
 We are indebted to the referees for their invaluable feedback, which served to enhance the quality of the paper. We made use of NASA's Astrophysics Data System as well as the VizieR and Simbad databases at CDS, Strasbourg, France and data from the European Space Agency (ESA) mission \emph{Gaia}\footnote{https://www.cosmos.esa.int/gaia}, processed by the \emph{Gaia} Data Processing and Analysis Consortium (DPAC)\footnote{https://www.cosmos.esa.int/web/gaia/dpac/consortium}. Funding for DPAC has been provided by national institutions, in particular, the institutions participating in the \emph{Gaia} Multilateral Agreement.

\bibliography{refs}
\clearpage
\end{document}